# THE WINDS, METALLICITY, ROTATIONAL VELOCITIES AND MASS LOSS RATE OF THE HOT, CONTACT BINARY TU MUSCAE (HD100213)

Raymond J. Pfeiffer
Professor Emeritus of Physics and Astronomy
*Department of Physics*
*The College of New Jersey*
*Ewing, NJ 08628 USA*
*Pfeiffer@tcnj.edu*

Abstract

A model of the TU Muscae binary system has been developed by a study of 23 *SWP* spectrophotometric images obtained with the *IUE* satellite telescope and downloaded from the *IUE* Archive.. The images are well distributed in Keplerian orbital phase thereby permitting a simultaneous fitting of the C IV wind-line profile by the SEI method and the light curve for the same bandpass by means of a program similar to that of Wilson and Devinney. The result is a set of parameters characterizing the physical and geometric properties of the wind envelopes surrounding the stars.

Surprisingly, there is no evidence for a P Cygni profile or strong, distinguishable shock front in the system, as has been found for similar investigations of EM Carinae and HD159176. This is probably a result of the contact nature of the binary and the high temperature environment of such a shock. That is, most of the carbon ions in the shock are more highly ionized. Based on the parameters for the *SEI* fit to the C IV profile, the value for the ionization fraction of C IV in the wind was calculated to be $10^{-4}$. With this value, the mass loss rate, $\dot{M}$, calculated from two independent equations, was found to be about $10^{-6}$ $M_\odot$/yr.

The line blanketing or metallicity was found to be erratically variable with orbital phase and time, indicating a variable amount of fast moving, dense clouds in the winds and/or a great amount of turbulence. The meaning of rotational velocities for the stars is problematic and depends on what point on the photospheres one is considering.

## 1. Introduction

A series of 2 previous papers has developed models of the wind structure and activity for four massive, close binaries. The first of these concerned the "well-separated" pairs Y Cyg and CW Cep (Koch, *et al.,*1996), while the remaining one studied in that paper, HD159176, is a much tighter but not a contact system. The second paper dealt with EM Car (Pfeiffer, *et al,*, 2004). The current contribution looks at TU Mus, a true contact (or double contact in some notation) object in a similar way. *A priori*, one cannot tell if the contact nature of the object will change the winds from the character known for the 4 binaries already studied.

A brief history for TU Mus has been summarized by Stickland *et al*. (1995), by Terrell *et al.* (2003), who note the importance of this very hot star, and most recently by Penny *et al*. (2008). The first two studies conclude with the confounding result that a high-weight, UV radial velocity curve from *IUE* images yields a mass ratio, q, which apparently disagrees significantly with the mass ratio from an adequate "blue" radial velocity curve (Anderson *et al.*, 1989) supplemented with 2 normal points from Terrell *et al*. This matter has been discussed in detail by Terrell *et al*. and they conclude that the source of the disagreement is an enigma. However, the seeming difference between the mass ratios from the two latter observational sets is not significant when realistic errors are imposed on the parameters. The most recent radial velocity analysis by Penny *et al*. (2008) concludes that the UV radial velocity analysis yields the correct value for q.

The study by Stickland *et al.* (1995) established precise orbital dimensions ($a\ sin(i) = 15.41 R_\odot$) and masses $M_1 \sin^3(i) = 15.70 M_\odot$ and $M_2 \sin^3(i) = 9.85 M_\odot$, yielding a mass ratio q = 0.627. A light curve analysis by Anderson, *et al*. (1975) yielded i = 76°. However, the study by Terrell *et al*. (2003) determined a value of i = 77.8° and a mass ratio of 0.651, which, as mentioned above, is the same number obtained by Anderson, *et al*. within $2\sigma$. The value for q implied from the analysis buy Penny *et al*. (2008) is 0.625.

The IUE archives hold 23 high quality SWP images for the contact binary system TU Muscae. The aim of the present paper is to (1) model the winds and their interactions in this binary by means of the same *IUE* images used by Stickland *et al.* (1995) for their velocity analysis, (2) study the variation of the metallicities or line blanketing in the spectra, (3) address the matter of the rotational velocities of the stars and (4) calculate the mass loss rate for the system that results from the winds.

For this binary, the Si IV and N V doublets within the *SWP* bandpass are too weak to provide useful information. Hence, this paper is limited to an analysis of the C IV feature near 1548 Å for the winds. This profile has been fitted with a program based on the *SEI* program developed by Lamers *et al*. (1987), which was first adapted for use with binary stars by Pfeiffer, *et al*. (1994) and was referred to as *BSEI*. For this study, substantial improvements have been to the *BSEI* program for fitting the UV line profiles. These improvements take into consideration the variable line blanketing that its observed from image to image. The results of this study have also made it possible to calculate the ionization fraction, $q_i$, for the amount of C IV in the wind and a determination of the mass loss rate for the system.

Also, an analysis of the Fe II, He II and C I features near 1640 Å was made to determine the rotational velocities of the stars. But, the concept of rotational velocities for the stars in a co-rotational, over-contact system is ambiguous and such a determination is fraught with problems. As a result, there are no unique answers.

## 2. The Mass Ratio

The following discussion on the mass ratio was written mostly by the late R. H. Koch (Koch, R. H., 2001) and was originally intended to be part of this paper:

The mass ratio for the TU Mus system has been a matter of contention in the literature. The situation may be summarized in this way. The visible-band photometry and spectroscopy essentially converge on q = 0.651, while the UV continuum light curve is most consistent with the UV radial velocities leading to q = 0.625. At least formally, these two values do not overlap at $3\sigma$. There is, however, a way to understand and resolve this difference. Choose the two photospheric temperatures given by Wilson and Rafert, (1981). (Their procedure cannot really determine these temperatures with high accuracy but the ratio of the temperatures is significant.) These temperatures lead to theoretical flux levels in the model atmospheres of Kurucz (1994) at any tabulated wavelength. For the UV continuum the ratio of the flux levels for the two temperatures is 1.58 while the same ratio is 1.31 in the visible.

It seemed useful, therefore, to see if the UV light curve contained any information which would bear on the confused mass ratio. Bradstreet's Program, Binary Maker 2 (Bradstreet, 1993) was used to calculate theoretical light curves which preserved the parameters of Wilson and Rafert (1981) but used UV limb darkening coefficients extrapolated from the calculations of van Hamme (1993).

This same program was then used to calculate a second theoretical light curve which used the UV mass ratio and preserved the contact configuration by changing to the appropriate gravitational potential. The difference between these two theoretical curves is not great, but the second one with the mass ratio developed from the UV cross-correlation procedure was slightly to be preferred. One can say this in a somewhat different way: the UV continuum light curve gives no reason to prefer the visible-band mass ratio. It must also be noted that the fit to the 0.25 - 0.75 phase interval is not so good as for the other half of the period. The sense of the problem is that the observed data remain slightly bright compared to the theoretical light curve. A problem not understood.

With the component stars in contact, the absorption and re-radiation of one star's flux by another and the differential between the two processes (distilled into the astronomical colloquialism of "the reflection effect") is very important. It displaces the light centers of the stars away from the positions of their projected mass centers toward the systemic mass center. The effect on the cooler, fainter star is the more important because it causes that star to seem to move in a smaller orbit in the UV than in the visible. The process does logically affect both the UV continuum and the UV line centroids and accounts for the consistency of the UV spectroscopy and light curve. The recognition of the effect is quite old. Kuiper first presented precepts for correcting it in 1938. However, the message here is somewhat different than the classical understanding. The latter may be interpreted to mean that the visible-band mass ratio is more reliable than the UV one, even though the latter is derived by a procedure which possesses the highest internal weight. This conclusion also negates some of the force of the dictum that the UV, where most of hot star radiation is emitted, is the proper electromagnetic domain to study hot stars  This strongly indicate that the mass ratio given by the UV analysis is the correct one.

### 3. Fitting the UV Light Curves

For HD 159176, Pachoulakis (1996) showed that the UV Spectrophotometry in an *IUE* SWP bandpass yields a good light curve, even if the flux modulation is only that of an ellipsoidal binary light variation and that even these modest light curves can be modeled usefully. The UV light curves for Y Cyg and CW Cep had previously been found to resemble the visible-band ones, if the resonance wind lines were avoided in selecting the bandpasses. Consequently, two light curves for TU Mus were extracted from the *IUE* spectra by integrating the observed flux over two bandpasses. The first is a continuum bandpass from 1450 to 1490 Å, and the second encompasses all of the C IV profile from 1530 to 1560 Å. A 200 Å wide bandpass between 1600 and 1800 Å was also investigated to see if an improvement in the noise in the continuum light curve could be obtained. It could not.

A study of these light curves shows that the UV radiation of TU Mus is intrinsically variable, aside from the usually binary induced causes. The first 4 images lead to continuum flux levels which are consistently brighter by 10% than the levels for all later images. The earlier and later sets of images are separated by more than 2,400 days or 1,700 cycles. The flux difference seems to have no effect on the Doppler displacements of the photospheric lines in the study by Stickland *et al.* (1995) and it has no known cause at present. However, it may be related to the highly variable metallicity and transient emission clouds that has been discovered for the system in this paper.

The UV continuum light curve may be faced against the visible-band ones already analyzed by Anderson, *et al.* (1989) and Terrell *et al.* (2003).   This has merit even though the IUE light curve is populated by only 19 data points (the four earlier *IUE* images having been excluded). The resemblance between the three data sets is generally quite strong.

The light curve for the C IV feature has very little information in it. Although morphologically the same as the continuum one, it suffers from shot noise in the spectrum because the line is so strong. A reasonably good fit to the UV continuum light curve and *u* bandpass of Terrell *et al.* (1998) has been obtained using a program, *UVLCP*, developed by employing techniques similar to that utilized by Wilson and Devinney (1971) and Wilson (1976). See Fig. 1 below.

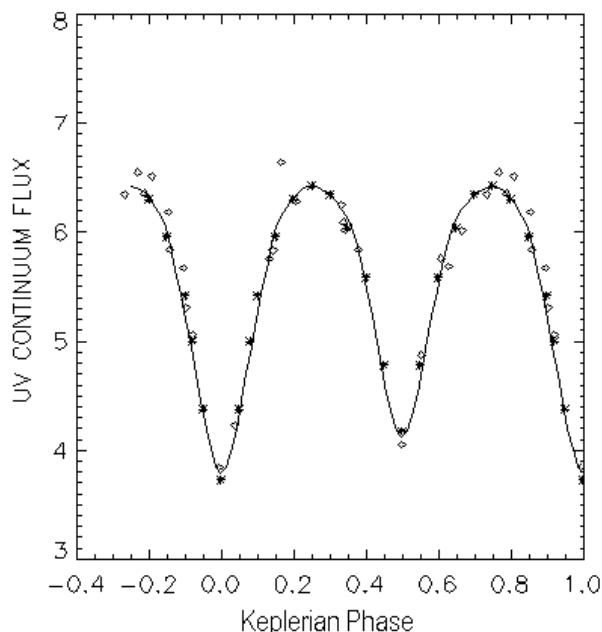

Fig. 1 displays three continuum light curves for TU MUS. The diamonds represent observations derived from the *IUE* blanketed continuum (1450 - 1490 Å) outside the C IV line profile. The stars are data points taken from Terrell et al's published light curve in the u bandpass, renormalized to the IUE flux levels, and the solid curve is a fit using a program developed by the author following a methodology similar to that published by Wilson and Devinney (1971).

In this program, the light assigned to each star utilizes the light ratio known from the photometric solution of Terrell at al. (2003). The out-of-eclipse, fiducial flux level of the system was determined from the image taken at phase 0.735 for each bandpass. This flux was then partitioned among the stars, wind envelopes, and a possible shock. The partitioning parameters are input parameters for the program and are varied to achieve a good fit. For a given set of input parameters, the light curve was generated taking into account all binary interaction effects: (1) all ellipsoidal variations contributing to the systemic light, (2) photosphere-photosphere eclipses, (3) photosphere-wind envelope eclipses, (4) photosphere-shock eclipses, (5) wind attenuation of the photospheres, shock, and other wind, and (6) shock attenuation of the photospheres and wind envelopes.

The primary purpose of achieving such a fit is to test the algorithms so developed that, when they are integrated into the binary interaction program *BSEI* (Pfeiffer and Stickland, 2004), the latter may be confidently brought to bear upon fitting the C IV wind-line profile. The binary system, as represented by these algorithms, is depicted in Fig. 2. This diagram was generated using the graphics available in *IDL* software installed on a SUN Blade 150 workstation and using a program written by the author employing techniques similar to those given in Kallrath, J. & Milone, E. F. (1999). A mass ratio of 0.625 was used and the unit of distance is approximately 16R$_\odot$, based on values given in Penney et al. (2008) for $a_1 \sin(i)$.

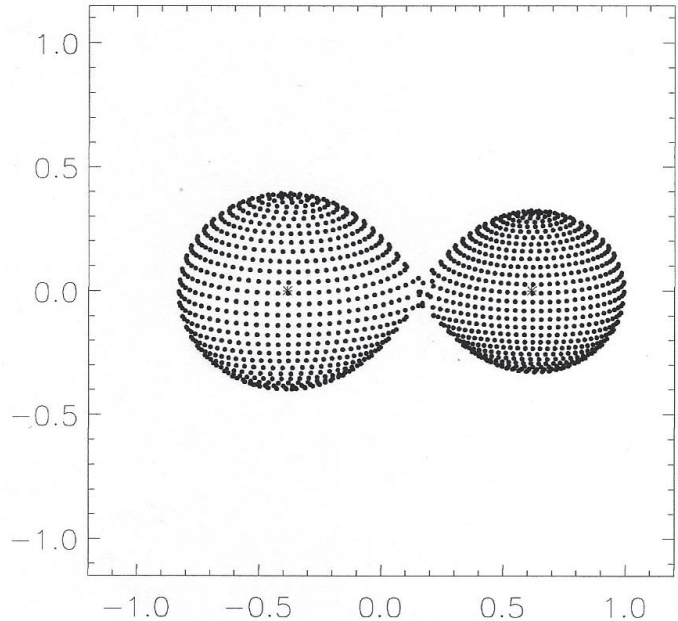

Fig. 2. The TU Muscae system as represented by the algorithms used to produce the light curve fit shown in Fig. 1. The barycenter is located at coordinates 0,0 and the mass ratio is 0.625. The centers of the two stars are denoted by asterisks. The scale is such that the center of the primary from the barycenter is 6.08 solar radii.

## 3. The *BSEILB* Fitting of the C IV Profile

The concern about the mass ratio is not an inconsequential one, for it determines the stellar dimensions in this contact system and thus locates the bases of the stellar winds. From the arguments presented in section 2, it was decided to use q = 0.625, although it isn't clear that this choice has to lead to the best wind geometry.

The winds can be modeled by fitting the profiles of the wind-sensitive lines. For TU Mus the N V doublet near 1240 Å is so severely convolved with the very broad interstellar $L_\alpha$ line, that there was no confidence in trying to isolate the stellar line. Also, the Si IV lines near 1400A are rather weak and give little indication of a wind. They appear to be dominated only by photospheric line absorptions. Consequently, all information about the systemic wind comes from the C IV doublet feature near 1548.2 Å. In addition, it was found that all 23 images can be used for the wind profile fitting. That is, in observed flux units, the 4 earliest images give line strengths and profiles which are indistinguishable from those of the later images. There is no current explanation why the continuum and line fluxes should be uncoupled in this way.

To begin, the *SEI* Program developed by Lamers *et al.* has been modified by the author to be applicable to binary star systems and has heretofore been referred to as *BSEI* (Pfeiffer et al. 1994). However, the author has further improved the latter program by taking into account the line blanketing. This has been done by the addition of Gaussian profiles to fit easily recognized line profiles. The very narrow ones being interstellar absorptions and the broader ones being interstellar absorption confluences or the infamous DACs (Discrete Absorption Clouds) mentioned by others, such as Howarth, *et al.* (1989). Other profiles may be actual photospheric metallicity blanketing. An attempt to identify the latter was made by inspecting the spectra to see if any of these features walked in wavelength from image to image in agreement with the changing radial velocities of the stars. However, this is difficult to do, as the many lines are walking back and forth and overlapping with one another from image to image. For this paper, the further modified program is referred to as *BSEILB*. This program was then employed to fit the C IV profile for the images at photometric phases 0.553 and 0.229. The table below lists the line blanketing features for the fit at phase 0.533. They are somewhat different at phase 0.229. No attempt was made to identify all the lines. They are a mixture of interstellar lines, DACs and photospheric lines. Depth is the fraction of the continuum level absorbed at the center of the line. The width is the half width in Å at half depth.

| WL (Å) | Depth | Width |
|---|---|---|
| 1533.30 | 0.46 | 1.2 |
| 1541.50 | 0.27 | 0.28 |
| 1543.70 | 0.15 | 0.60 |
| 1545.30 | 0.10 | 0.30 |
| 1549.00 CIV | 0.22 | 0.20 |
| 1551.42 CIV | 0.24 | 0.20 |
| 1554.25 | 0.14 | 0.22 |
| 1556.82 | 0.25 | 0.30 |

Photometric orbital phases for the IUE images used in the current paper have been obtained by converting the spectroscopic phases listed in Table 1 of Penny et al. (2008). The original SEI fitting procedure uses normalized flux levels. But this can be misleading and may indicate a stronger P Cygni profile than what actually exists. So, in this paper the normalized fluxes emerging from *BSEILB* were converted to *IUE* absolute fluxes. This was accomplished, by using the *IUE* blanketed flux levels between 1500 Å and 1600 Å, found from the dereddened IUE image, to define the slope of a linear interpolation over this wavelength interval. The continuum level was set sufficiently high so that blanketing features were real and allowing for the noise in the *IUE* fluxes.

The first fit is shown in Fig. 3, which clearly indicates a weak or non-existent P Cygni profile.

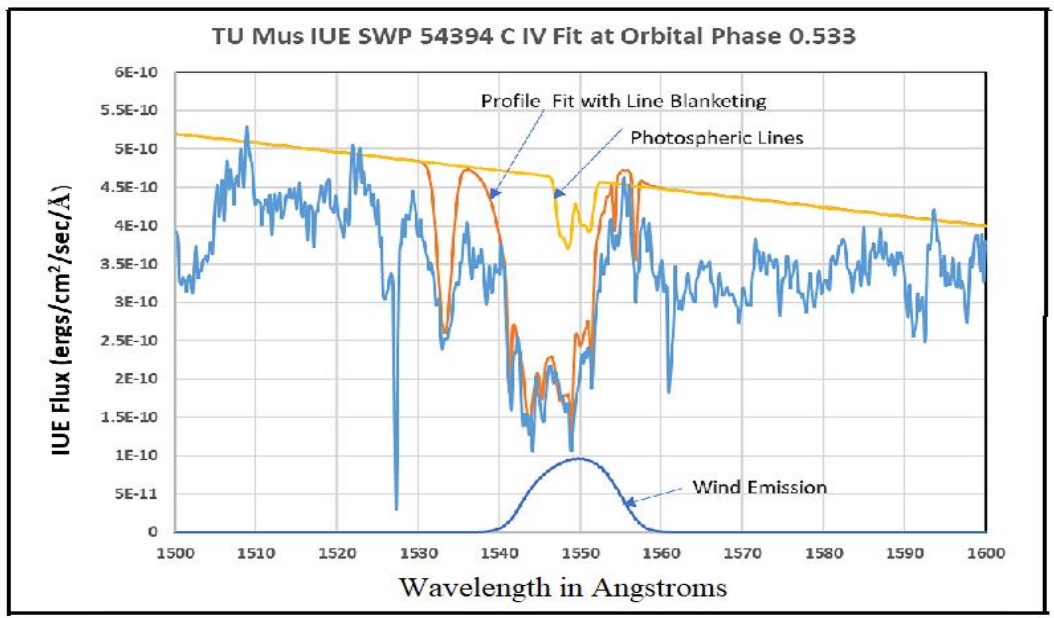

Fig. 3. The BSEILB fit to the C IV feature at 1548 Å

The low amplitude feature at the bottom of Fig. 3 separately displays the total wind emission, which has been modified to allow for eclipsing effects of a wind halo by the other star. The amount of such emission is critical for a best fit. The line features displayed to be hanging down from the continuum flux level are the two photospheric lines of the C IV doublet for each star. The photospheric lines have been modeled taking into account the eclipsing effects of the secondary by the primary at this phase. The proper strength of these lines, which are usually ignored in wind studies, make it possible to achieve the fit shown.

The total fit had to match an unblanketed continuum while taking into account the shot noise both within the line profile and in the adjacent continuum between 1500 and 1600 Å. In this way, one obtains optical depths that are more accurate for the C IV ion. The two interstellar C IV lines are easily recognized and fitted. These have total widths of about 1Å. In addition, there appear to be several other narrow lines (< 2.0 Å in total width) within the bandpass, which cannot be of photospheric origin. The dictum is that a photospheric line must be rotationally broadened by about 3.3 Å in total width at the continuum as judged from the He II feature at 1640.4 Å, as shown in Fig. 5 below. Such lines cannot conspire to produce narrow lines. However, several close interstellar lines can conspire to produce a broadened feature. Also, the question is, which of these features are real photospheric line blanketing profiles and which are not. The answer to this was decided by comparing all the images to see which narrow features were consistently present. It was decided that if a narrow feature appeared to be at the same wavelength with the same approximate strength in three or more images, it is a real interstellar line. In some cases, an interstellar line may be obliterated by the Doppler walking of unknown photospheric lines. On the other hand, some features may be transient DACs. If the feature seemed to indicate an Increasing blue Doppler shift with Keplerian phase, the probability is that the feature is a DAC.

There are two narrow lines near 1549.4 Å and 1552.0 Å that are separated by 2.6 Å, which is the nominal separation of the blue and red components of the C IV doublet. These obviously originate from an interstellar cloud that is red shifted relative to the telescope by 0.73 Å, since the IUE images have been corrected for the motion of the satellite. Adding some of the line blanketing and interstellar absorption lines great facilitates obtaining a best fit to the profile. The departure of the fit from the spectrum approaching terminal velocity at wavelength 1536 Å is a result of numerous, strong line blanketing features that were not modeled. But the sharp decrease in absorption near 1541Å is well fitted and this was a determining factor for the input parameters.

The fact that the C IV spectral feature shows no noticeable P Cygni profile at this phase is reinforced in the IUE image SWP54407, shown as Fig. 4. The latter was made at orbital phase 0.229, when a possible shock and the wind envelopes are minimally eclipsed by the stars and present the best profile seen projected in the plane of the sky.

All of this may be interpreted to mean that the number density of C IV ions is low in the shock front and in the halos of the winds surrounding the stars. This probably results from the thermalization of the kinetic energy of the wind by collision that causes the ionization of the carbon ions to higher states of ionization. In fact, the work of Stevens, et al. (1992) indicates the temperature in a shock may exceed $10^7$ K. It may also be possible that the high intensity ultraviolet radiation from the stellar photospheres contributes to the high ionization.

The photospheric absorption lines are relatively strong and contribute significantly to the absorption observed in the profile and indeed this doublet for both stars was modeled this way to achieve the best fit. The photospheric lines should be slightly weaker in Fig. 4, but left this way to show how important they are for a best fit. The photospheric lines are weaker at phase 0.229 than at phase 0.533. (Compare the Aphot values in Tables 1 & 2.) This may be attributed to an effective increase in the density and extent of the reversing layers of the stars at the latter phase. This is the equivalent to saying that there are DACs that happened to be near the vicinity of the radial velocity shifted wavelengths of the photospheric lines.

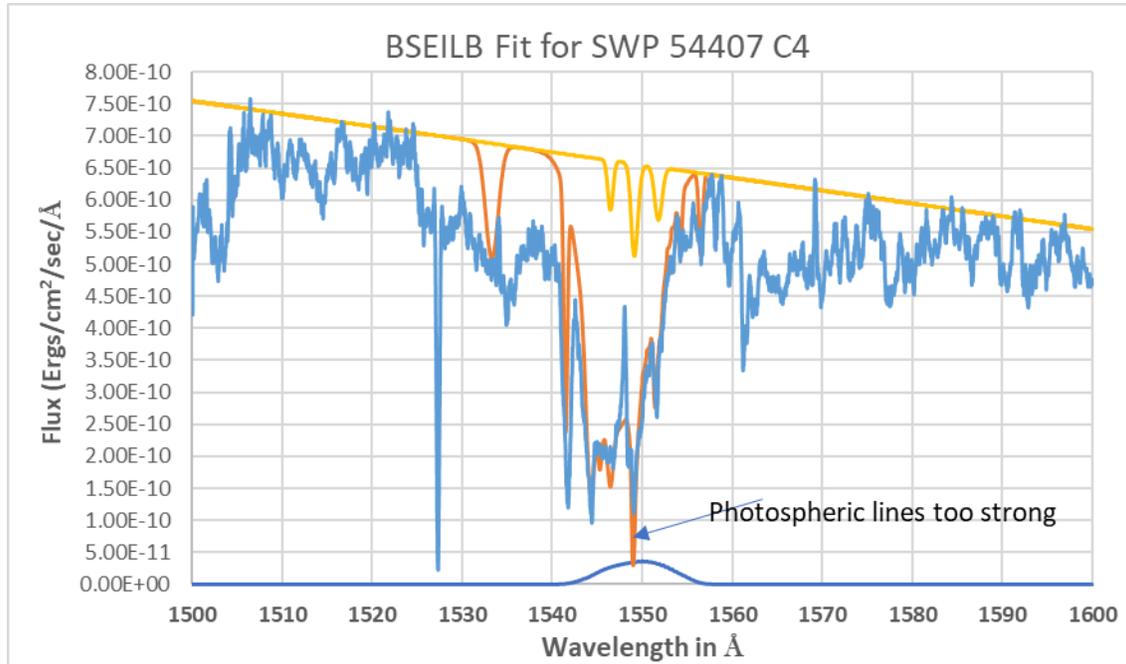

Fig. 4. This IUE image made at phase 0.229 also shows no evidence of a classical P Cygni profile, which should be clearly present here. The photospheric lines are purposely shown to be too strong in order to show how important they are for a best fit.

Table 1 presents the BSEILB parameters determined for the fit shown in Fig. 3 and Table 2 shows the parameters for the fit to SWP 54407. Note that DTV, and, hence the terminal wind velocity, is smaller at the latter phase. Also, much more wind emission is eclipsed at phase 0.229, as it should be, since each star is now seen side by side in the plane of the sky, as depicted in Fig.2, rather than seen almost as one in front of the other at a conjunction.

Essentially, most of the parameter nomenclature in the tables is that originally used by Lamars in the original *FORTRAN* code used for the *SEI* Program and is as follows:

$W_0$ is the distance from the star, in stellar radii, at which the opacity of the wind begins.
W1 is the distance from the stellar photosphere, in stellar radii, at which the terminal velocity of the wind is reached and the opacity vanishes
WGauss is the width of the assumed Gaussian turbulent velocity distribution.
TTOTB & TTOTR are the total integrated optical depths through the wind for the blue and red wavelength components of the ion.
Alfa0 and alfa1 are parameters for determining the variation of the opacity, κρ, through the wind.
DTV is the wavelength difference between the blue component rest wavelength and the wavelength where the terminal velocity of the wind occurs.
Gamma is an exponent on the law for the variation of the opacity of the wind from the star.
AphotB and AphotR are the depths of the photospheric lines for the ion in normalized flux units.
Wphots are photospheric line widths in *SEI* normalized units, not Angstroms..
Lf is the light fraction of the system contributed by each of the stars.
FWE is the fraction of the wind emission that remains after being eclipsed by the other star.

Table 1
BSEILB Parameters for SWP 54394 C IV at Phase 0.533

| Parameter | Primary Star Wind | Secondary Star Wind |
|---|---|---|
| $W_0$ | 0.01 | 0.01 |
| W1 | 1.0 | 1.0 |
| WGauss | 0.28 | 0.28 |
| TTOTB | 2.50 | 0.50 |
| TTOTR | 1.20 | 0.21 |
| Alfa0 | 0.05 | 0.01 |
| Alfa1 | 7.0 | 7.0 |
| DTV | 12.05 | 12.05 |
| Gamma | 0.50 | 0.50 |

| Parameter | Primary Star | Secondary Star |
|---|---|---|
| APhotB | 0.42 | 0.41 |
| AphotR | 0.29 | 0.31 |
| WphotB | 0.05 | 0.05 |
| WphotR | 0.05 | 0.05 |
| Lf | 0.55 | 0.45 |
| WFR | 0.75 | 0.64 |

Table 2
BSEILB Parameters for SWP 54407 for C IV at Phase 0.229

| Parameter | Primary Star Wind | Secondary Star Wind |
|---|---|---|
| $W_0$ | 0.01 | 0.01 |
| W1 | 1.0 | 1.0 |
| WGauss | 0.28 | 0.28 |
| TTOTB | 2.50 | 1.00 |
| TTOTR | 1.20 | 1.50 |
| Alfa0 | 0.05 | 0.01 |
| Alfa1 | 7.0 | 7.0 |
| DTV | 8.0 | 8.0 |
| Gamma | 0.50 | 0.50 |

| Parameter | Primary Star | Secondary Star |
|---|---|---|
| APhotB | 0.30 | 0.30 |
| AphotR | 0.20 | 0.20 |
| WphotB | 0.05 | 0.05 |
| WphotR | 0.05 | 0.05 |
| Lf | 0.55 | 0.45 |
| WFR | 0.27 | 0.25 |

Surveying all the IUE images of TU Mus, one sees a variety of changes from phase to phase, some of which are phase locked and others that are transients. Particularly note the emission spike in Fig. 4 near 1547 Å, which is a transient and not seen in any if the other images. Since it is blue shifted relative to the rest wavelength of C IV, it is emission from a dense cloud moving towards the telescope, but to the side of one of the stars. It is the reverse of a DAC.

The last parameter is needed because the fluxes emanating from the system for each star are calculated separately, assuming the other star does not exist. For example, 75% of the emission from the wind of the primary star remains in Fig. 3, after being eclipsed by the secondary star at the orbital phase of the system for which the fit was made. The value found for DTV to be 12.05 Å for both stars corresponds to a terminal velocity of 2335 km/sec. The value of DTV in Table 2 corresponds to a terminal velocity equal to 1550 km/sec. The uncertainty in DTV was just about 0.05 Å, corresponding to an uncertainty in the terminal velocity of ± 10 km/sec. Somehow, the winds from both stars moving along the same line of sight toward the system may cause the difference in the terminal velocity.

Errors of the parameters are about one or two units in the last decimal pace, but such errors were difficult to determine. This is because a best fit was determined by varying all the parameters by trial and error. Some parameters were more sensitive for a best fit than others.

The determination of the *BSEILB* denormalization constants can be a source of error for determining the continuum level and its slope. In addition, other errors may be introduced by the assumption that the fluxes between 1500 and 1600 may be fitted by simply a linear interpolation. Never-the-less, the Fits shown in Figs. 3 and 4 appear to be remarkably good.

## 4. Metallicity and Mapping Enhancements in the Density of the Winds

The spectrum of a star shows many spectral lines of various depths, or strengths, that overlap and often form confluences. The absorption of flux that these lines cause is referred to in spectroscopists' jargon as "blanketing". The metallicity of a stellar atmosphere, z, is defined as the fraction of star's integrated unblanketed continuum flux that is absorbed by the many metal ions in a star's atmosphere. The absorption is commonly referred to as taking place in the reversing layer, which is the upper most layer of a star's photosphere.

A casual inspection of stellar spectra shows that z varies considerable over the spectrum, aside from a wind profile, such as, C IV. Therefore, it is difficult to specify exactly what z is for any broad wavelength band. The question is what should z be for the observed spectrum of a binary star, since it is a composite of two stars. Furthermore, each star exudes a wind, and these winds collide and form a turbulent shock front and streams that wrap around the other star. The shock and turbulent streams in the TU Mus system probably look similar to the hydrodynamic simulations for colliding winds in OB binary stars calculated by Stevens *et al.* (1992), as shown in Figs. 12 and 15 of their paper.

All and all, the total, or integrated amount of absorbed flux, **TIAF**, in a spectral bandpass caused by the blanketing should remain constant, unless the wind from the companion star alters it in some way and the changing geometric and photometric changes that result from the revolution of the stars about their barycenter also induce a change. The changes in the metallicity is essentially the result of an enhancement of each star's reversing layer. This is similar to the absorption seen in the C IV profile which increases the depth of the C IV profile in addition to that caused by photospheric lines.

The question remains whether the presence of a shock front, clouds or wrap around wind streams effect the metallicity. To answer this question, the value of TIAF was determined for different *IUE* images made at different orbital phases. The bandpass selected was 1500 to 1600 Å, excluding the wavelength interval containing the C IV feature, 1540 – 1552 Å. The results are presented in Table 3 below. Integrated *IUE* flux units over a bandpass are ergs/cm$^2$/sec. Typical shot noise in the flux of the TU Mus spectra is +/- 0.07e-10 flux units. An integration of the flux over a wide bandpass would greatly reduce this error so that uncertainty would be in the 3rd decimal place of the values in Table. 3. The changing value of TIAF rather than z is the best indication of how the wrap around wind streams and clouds affect the metallicity and allow for a mapping of this material in the system. Notice that an increase in TIAF is not always associated with an increase in z. **TIUBF** is the total integrated unblanketed flux, that is, the supposed unblanketed continuum that was fitted to the spectrum, as shown in Figs. 3 through 7.

The results for TIAF are also shown in Fig. 8 below. The curve is somewhat erratic with sudden jumps, indicating that the wind is lumpy, consisting of various clouds that may come and go, or turbulence. The clouds may be imagined to be passing in front of the stellar photospheres of the stars as the stars revolve in orbit and the clouds may have independent motions of their own as the winds flow outwards. The decrease in metallicity at phase 0.49 may indicate the eclipse of a dense cloud or shock between the stars.

The glitch between phases 0.826 and 0.836 occurred over an interval of 1993 days. The sudden drop at phase 0.833 is believed to be real within the errors of measurement and further indicates the turbulence that occurs in the wind. The increase in metallicity at phase 0.353 may be the result of a dense cloud passing in front of one of the stars. The image at phase 0.229 is separated from the image at 0.833 by 0.9 days with the image at phase 0.229 being earlier by 0.6 in phase. So, the cloud may have persisted for this long but has changed its geometry and/or density. All of this variability of the line blanketing is expected in light of the work of Stevens, *et al.* (1992). The latter have shown that a shock would consist of various turbulent eddies that move within and along the front and that these eddies would act as clouds.

Table 3

| SWP | Phase | TIUBF | TIAF | z |
|---|---|---|---|---|
| 37879 | 0.045 | 1.67e-6 | 3.82e-7 | 0.23 |
| 37880 | 0.100 | 1.86e-6 | 3.68e-7 | 0.20 |
| 54382 | 0.122 | 1.76e-6 | 2.85e-7 | 0.16 |
| 54353 | 0.158 | 1.90e-6 | 3.29e-7 | 0.17 |
| 54407 | 0.229 | 1.99e-6 | 3.12e-7 | 0.16 |
| 54356 | 0.283 | 1.93e-6 | 3.05e-7 | 0.16 |
| 19642 | 0.301 | 2.13e-6 | 4.60e-7 | 0.21 |
| 19643 | 0.346 | 2.35e-5 | 3.93e-7 | 0.17 |
| 54411 | 0.353 | 2.34e-6 | 5.64e-7 | 0.24 |
| 15916 | 0.387 | 1.86e-6 | 3.89e-7 | 0.29 |
| 54390 | 0.399 | 1.69e-6 | 3.00e-7 | 0.18 |
| 54360 | 0.416 | 1.73e-6 | 4.05e-7 | 0.23 |
| 54415 | 0.490 | 1.51e-6 | 2.16e-7 | 0.18 |
| 54394 | 0.533 | 1.39e-6 | 2.90e-7 | 0.21 |
| 19651 | 0.657 | 2.15e-6 | 4.32e-7 | 0.20 |
| 54371 | 0.702 | 2.07e-6 | 4.18e-7 | 0.20 |
| 37906 | 0.826 | 2.05e-6 | 3.87e-7 | 0.19 |
| 54426 | 0.833 | 1.93e-6 | 3.46e-7 | 0.18 |
| 54374 | 0.836 | 1.96e-6 | 3.87e-7 | 0.20 |
| 37907 | 0.872 | 2.00e-5 | 4.28e-7 | 0.22 |
| 54378 | 0.992 | 1.36e-6 | 3.19e-7 | 0.24 |

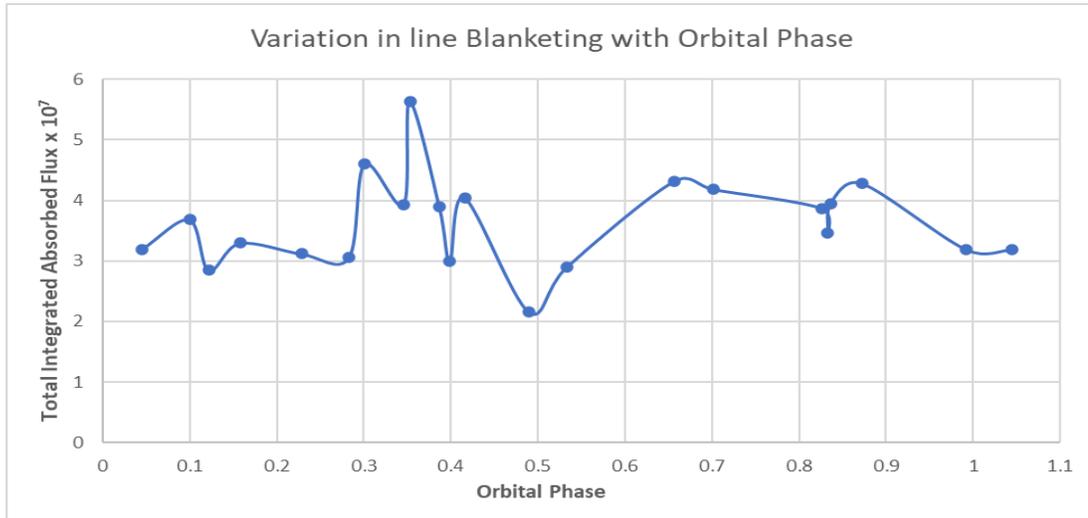

Fig. 5. The variation of the amount of TIAF (Line Blanketing), with photometric orbital phase, that results from the changing density aspects of the wind as seen projected onto the photospheres of the stars. Errors are believed to be less than +/- 0.01e-7 flux units.

## 6. The Binary Stars and Wind Interactions

The program used to fit the light curve of the C IV profile is the same one simultaneously employed to fit the continuum light curves discussed in section 3. This imposes constraints on the input parameters of the program, thereby resulting in a self-consistent method of analyzing both data sets.

At the quadrature phases, the star and wind interactions demonstrated in the *IUE* images are at a minimum and the light curve is at a maximum. As the stars approach conjunction, the interactions among the stars and their wind envelopes affecting the line profiles should increase. This is because the primary star is eclipsing the secondary star's photosphere, wind envelope and a shock front between the stars. This means the depth of the wind-line changes relative to the fiducial continuum level. One might also expect that another effect will be a change in the absolute value of the metallic line blanketing, z. Such changes could occur because of: (1) an ellipsoidal variation in the stellar fluxes, (2) the presence of clouds or changes in the density of the wind and (3) photospheric eclipses. As was shown in section 5, the presence of clouds and/or a turbulent shock front is most likely the main cause the of the variation in the line blanketing, which is erratic.

At phase 0.229 (SWP54407), the dereddened continuum flux extrapolated to the center of the C IV profile is about 6.05e-10 *IUE* cgs units. This is very nearly the maximum value for such flux, which is mainly the result of an ellipsoidal variation, since the eclipses are very shallow.

## 5. Rotational Speeds or Velocities

The concept of stellar rotational velocities for an over-contact binary system such as TU Mus is a problematic one. During the evolution of the system, the individual rotational velocities of the two stars, as they approached one another and came into contact, were subject to tidal forces that would result in those velocities being synchronized with the orbital motions of the stars around their common barycenter. Every point on the photospheres would then revolve about the barycenter with a velocity that depends on the distance of that point from the barycenter. Hence, points on the outer limbs of the photospheres of the stars would have a higher velocity than those points closer to the barycenter. This would result in asymmetric line profiles for the photospheric spectral lines that could not be fitted with simple Gaussians. In Fig. 2, it is apparent that the outer limb of the less massive star is farther from the barycenter than the outer limb of the more massive star and therefore should display a greater velocity.

With this in mind, an attempt to determine the rotational speed of the outer limb of both stars was made anyway, by the simultaneous fitting of Gaussians to the Fe II feature at 1631.13 Å, the He II profile at 1640.37 Å and the C I profile at 1657.38 Å. This was first done at phase 0.490, that is, near secondary conjunction, using a separate program written by the author. The result is displayed in Fig. 6. Orbital velocities of the center of mass of the stars were taken from Penney et al. (2008), though these had to be changed slightly to achieve a best fit. Values for the wavelengths were taken from the online NIST ADS Catalogue. The Fe II and C I profiles are contaminated by other lines, but the core of the He II feature looks rather pristine. There is no indication in the He II profile that He II ions are entrained in the winds in the system. The deep absorption spike below the C I feature is probably a strong C I interstellar line that is slightly blue shifted relative to the TU Mus system.

The photospheric lines of both stars making up the total profile fit (the red Gaussians) are shown separately. The lines of Star 1 are the blue Gaussians, while the lines of star 2 are the gray Gaussians. By demanding a good fit to all three lines simultaneously puts constraints on determining the parameters for the fitting procedure, thereby allowing for a more accurate determination of the line widths. For example, wider Gaussians for the Fe II lines cannot be justified, unless one assumes that the Fe II lines are formed in a different part of the photospheres of the stars than the other lines, which probably is unlikely. In any event, Doppler rotational broadening should be the dominant factor for determining photospheric line widths. Stark broadening would only result in a wider, Voight profile.

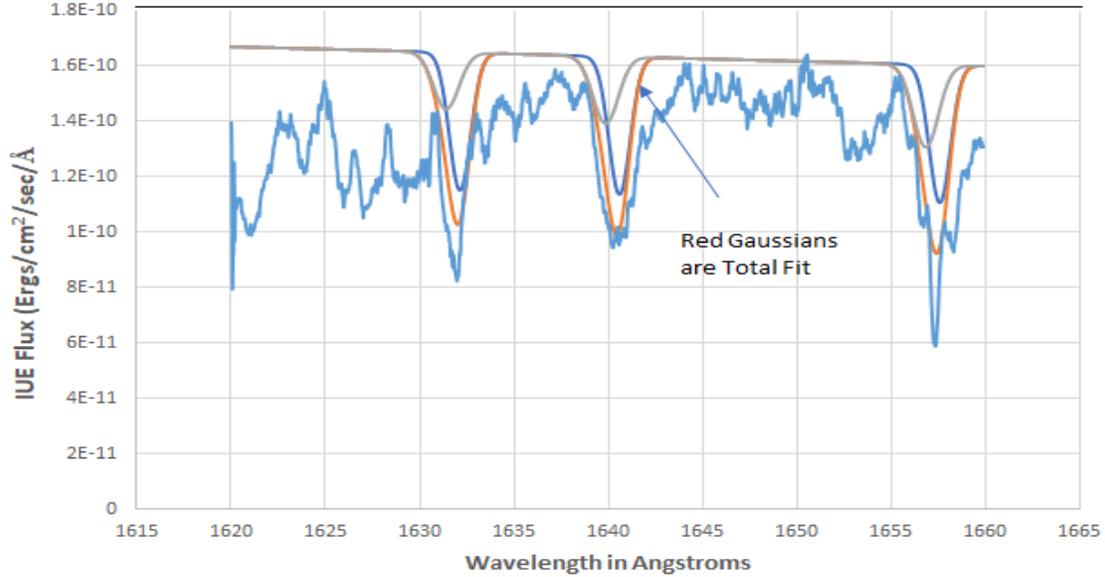

Fig. 6. Gaussian profile fits for Fe II, He II & C I absorption line features in *IUE* SWP 54415 made at photometric orbital phase 0.490. The primary star's lines are blue and the secondary's are gray. The red Gaussian is the total fit.

The half-widths of the Gaussian profiles at the continuum are 1.62 Å for the stars. This corresponds to a value v sin(i) = 298 +/- 5 km/sec. This value is larger by about 10 km/sec than other methods of determining the rotational speeds, such as Penny et al. (2008) but is agreement with the value obtained by Terrell et al. (3003) of 290 km/sec and the value of 306 km/sec reported in the catalog of Bernacca & Perinotto (1970). However, as discussed above, the concept of individual rotational velocities for an obviously co-rotating system such as TU Mus is problematic.

The outer limb of the larger star must be co-rotating at a velocity greater than that of the center of mass of the larger star, orbital velocity, which, according to Penney et al. is 215 km/sec. This is in agreement with the rotational velocity for the stars to be 298 km/sec, though the limbs of the secondary should be moving at a greater velocity than the limbs of the primary.

The only way to make a precise determination of the rotational velocities is by measuring the half widths of profile fits at the continuum. Therefore, setting the continuum level is critical for determining the rotational velocities. Lowering the continuum would result in smaller rotation velocities and setting it higher would result in larger rotational velocities. The continuum must be properly identified taking into account the line blanketing of the spectrum and the shot noise in the flux calibration, which is about ± 0.07e-10 ergs/cm$^2$/sec/Å in Fig. 6. The line blanketing and other line confluences are strong to the short wavelength side of the F II feature compared with that between the He II and C I profiles. The former region of the spectrum is replete with many relatively strong, overlapping Fe II and Fe III lines. Lowering the continuum more could not be justified in light of these facts.

A fit to these same features in SWP 54407 at orbital phase 0.229 has also been carried out with the same parameters. The result is shown in Fig. 7.

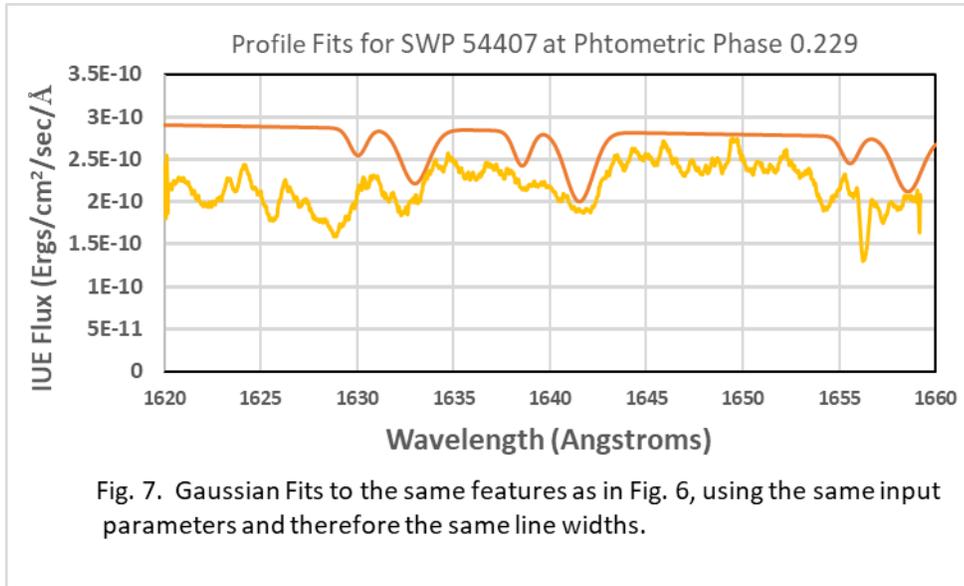

Fig. 7. Gaussian Fits to the same features as in Fig. 6, using the same input parameters and therefore the same line widths.

Obviously the line profiles have a totally different appearance at this phase. For the fits shown above, the photospheric line profiles for the two stars do not overlap and are seen as perfectly symmetric Gaussians. However, as mentioned above, they cannot be symmetrical profiles. Even though the individual stellar lines for these profiles are more separated, the picture is somewhat more complicated by the fact that all the other photospheric lines are overlapping and blending together. Moreover, these fits do not appear as good as those in Fig. 6.

Now the barycenter of the system is shifted from the inner LaGrangian point toward the center of the more massive star in accordance with the mass ratio. This is shown in Fig. 2. In this sense, the Gaussian profiles shown above are not correct, since the profiles of the two stars should be asymmetric. At this phase, there is less of the more massive star moving towards the observer than is moving away and more of the less massive star that is moving away. In this sense, Figure 3 of the H$\beta$ profile in Terrell *et al*. (2003) is more correct but somewhat misleading, for it gives the impression that the profiles of both stars overlap and are symmetrical. Therefore, a best fit to the same profiles shown above was carried out that resulted in line widths corresponding to a rotational velocity for both stars to be v sin(i) = 516 ± 15 km/sec! This fit is displayed in Fig. 8.

In this Figure, only the total fit is shown. Indeed, the fits to all three of the line features look better and somewhat reasonable than those in Fig. 7. In addition, the profiles do resemble those in Fig. 3 of Terrell *et al*. However, the profiles may be overfitted. The co-rotational velocity of the outer limb of the secondary, which is at a distance of 14R$_\odot$ from the barycenter, is 511 km/sec in the plane of the orbit and so v sin (i) should be less. In addition, at this phase, the rotational velocity of the outer limb of the secondary should be greater than for the primary.

The result of all of this is that there is no unique rotational velocity that can determined for the stars and that the assignment of a rotational velocity depends on what point of the photosphere of a star is being considered.

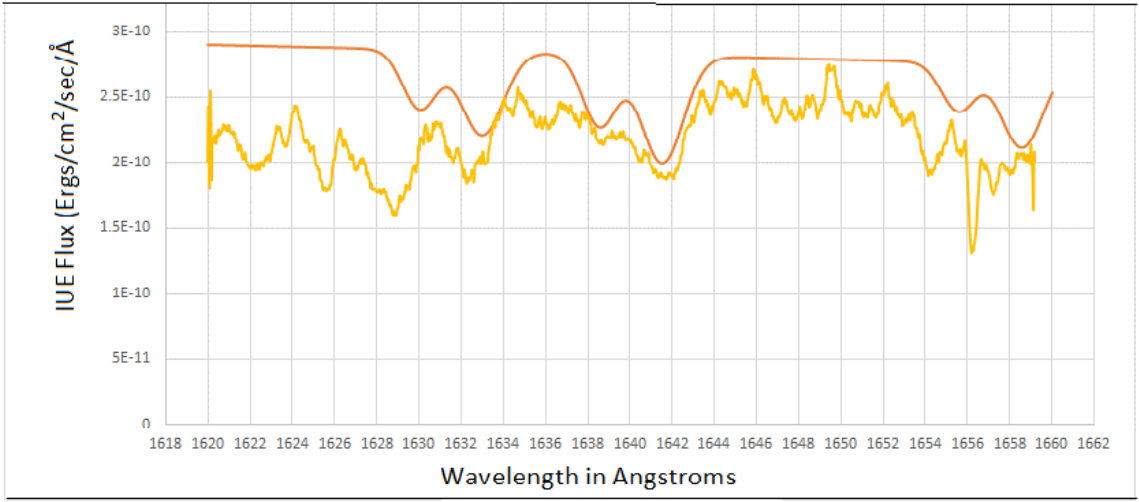
Fig. 8. Profile fits to the same images as in Fig. 6 but with much wider line widths to resemble overlapping symmetrical profiles.

## 7. The Ionization Fraction, $q_i$, for C IV

It is possible to calculate the value of $q_i$, if the optical depth of the wind at a distance from the photosphere is known. The optical depth $\tau_{\Delta\lambda}$ in bandpass $\Delta\lambda$ is given by:

$$\tau_{\Delta\lambda} = f \iint n(t)\sigma(\lambda)\phi(\lambda)d\lambda dt. \qquad (1)$$

Here, t is the length of a path through a medium, $n$ is the ion density, $\sigma(\lambda)$ is the scattering cross-section and $\phi(\lambda)$ is the broadening function for a spectral line. Since we are not interested in the details of the line profile, the broadening function may be replaced by dividing the result of the integration by the Doppler broadened width of the profile, $\Delta\lambda_D$. Furthermore, the scattering cross-section is constant for our bandpass, $\Delta\lambda$, and it will be assumed the ion density is constant along the path of length t as a first approxiation. It turns out that this assumption is justified, since the variation of $\tau_{\Delta\lambda}$ with t is known and mainly depends on the variation of $n$. With the above simplifications, (1) then becomes:

$$\tau_{\Delta\lambda} = f\, n t \sigma_{\Delta\lambda}/\Delta\lambda_D. \qquad (2)$$

It follows that the number density of C IV ions in the wind, $n$, at distance t may then be estimated from:

$$n = \tau_{\Delta\lambda} \Delta\lambda_D/(f\, t\, \sigma_{\Delta\lambda}) \qquad (3)$$

Now $\tau_{\Delta\lambda}$ is the value of the optical depth of the wind at the distance from the star, $x = r/R_\star$, which is parameterized in the SEI program by velocity w. Now, from Groenewegen & Lamars (1989), equation (5):

$$\tau(w) = (T/I)(w/w_1)^{a_0}\{1-(w/w_1)^{1/\beta}\}^{a_1} \qquad (4)$$

The value for (T/I) emerges from the BSEILB program and was found to be 8.897. The values of $\tau(w)$ at various values of w along path t were then computed from (4) using the parameters in Table 1. The results are displayed as Fig. 9. The total integrated optical depth along a radial path through the wind from the photosphere to where $\tau(w)$ vanishes is 2.50, which is TTOTB in Table 1.

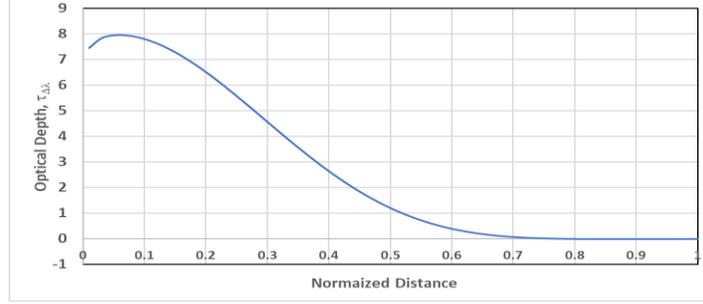

Fig. 9. Variation of the optical depth of the wind in TU Mus with distance from the photosphere. A distance 1.0 corresponds to 1 stellar radius, which is where the opacity of the wind was modeled to vanish.

The value of $\tau_{\Delta\lambda}$ was found to be 1.21 for w = 0.5. In addition, $f$ is the transition probability, 0.194. The value for $\Delta\lambda_D$ is taken from the value of DTV in Table 1 and is 24 Å. The parameter t is the distance along a line of sight extending from the surface of the star to the distance where the optical depth is evaluated. In this case, $t = 0.5R_\star = 3.6R_\odot$. The value $\sigma_{\Delta\lambda}$ is the scattering cross-section, given by $(\pi e^2/m_e c)\lambda_0^2/c$, which is 2.12e-20 cm$^3$ for the C IV profile. Hence, equation (3) yields a value for $n$ to be about 278 ions per cm$^3$.

The assumption that $n$ is constant over the path t, means the above value of $n$ is also the value at $0.5R_\star$. All things being equal, the ion density for the wind should follow the optical depth law, since equation (2) shows that optical depth mainly varies as $n$ varies. So, from Fig. 9, the value of the ion density near the photosphere should be about 6.4 times greater than at x = 0.5R. This is how one can justify removing $n$ from the integral over t in equation (1). The result is that $n$ is 1780 ions /cm$^3$ at the base of the wind.

Now, the value of the total mass density in a wind may be calculated as follows:

$$\rho_{wind} = (1 + \frac{n_{He}}{n_H} A_{He} + \frac{n_C}{n_H} A_C + \frac{n_N}{n_H} A_N + \frac{n_z}{n_H} A_z)n_H m_H \qquad (5)$$

Here $A$ is atomic weight, $m_H$ is the mass of the hydrogen atom (and $z$ represents all the other atoms other than helium, carbon, and nitrogen. Since the metal abundances are relatively small, they may all be represented by one term. Values for the abundance ratios relative to hydrogen were taken from Cox (2000) to be 0.090 for helium, 3.63 x 10$^{-4}$ for carbon and 8.80 x 10$^{-3}$ for nitrogen. For the other metals, which are partially ionized to some unknown degree, a weighted mean value has been taken to be 20.0 for $A_z$ and $n_z/n_H$ = 0.0001. The exact value is not significant, since and $n_z/n_H$ is so small. The sum of all the terms in the parentheses in (4) is then 1.498 and it follows that

$$\rho_{wind} = (1.498)[n_m/(n_r/n_H)]m_H. \qquad (6)$$

The number density $n_m$ is that measured for an atom in the wind and $n_r$ is the number density based on the abundance of that atom relative to hydrogen.

Typical wind densities, $\rho_w$, for hot stars at photospheric distances of $x/R_* = 1$ are about 10$^{-13}$ g/cm$^3$ (Lamers *et al.*, p. 251). Assuming $\rho_w = 1.0$ x 10$^{-13}$ g/cm$^3$ for TU Mus, equation (6) demands that $n_C$ be $1.45 \times 10^7$ ions per cm³. Hence, the number density found for C IV, is some small fraction of all the carbon atoms in the wind. Using the value for $n_{CIV}$ that was computed from (3), the fraction of C IV ions to the total number of carbon atoms is $1780 /1.45 \times 10^7 = 1.23 \times 10^{-4} \approx 10^{-4}$.

## 8. Mass Loss Rate

The mass loss rate may be calculated from equation (12) in Groenewegen and Lamers (1989), hereafter G&L, namely:

$$\dot{M} = 1.189 \times 10^{-18} [R_* v_\infty^2 \tau(w) x^2 w (dw/dx)][f \lambda_0 A_E q_i(w)]^{-1}$$

This was done using wind parameters from the BSEILB fit to the C IV profile. The parameters in the above equation are:

$R_*$ is the radius of the primary star in solar radii = 7.20 from Penney *et al*, (2008)
$v_\infty$ is the terminal velocity of the wind = 2335 km /sec.
$w$ is the speed normalized to $v_\infty$ and taken nominally to be 0.5 for the blue component of the doublet for use in the above equation.
$\tau(w)$ is the optical depth at speed $w$: $\tau(w)=(T/I)(w/w_1)^{a_0}\{1-(w/w_1)^{1/\beta}\}^{a_1} = 1.21$. T/I emerges from the BSEILB program to be 8.897. The parameters $a_0$, $a_1$ & $\beta$ = Gamma are given Table 1.
$x^2 = (r/R_*)^2 = (1.5)^2 = 2.25$. Here, r is radial distance from the stellar center along the line of sight.
$dw/dx = (1 - w_0)\beta(1- 1/x)^{(\beta-1)}(1/x^2) = 1.21$. See Appendix. Beta is the same as Gamma in Table I above.
$f$ is the transition probability = 0.194. (Taken from G&L Table III)
$\lambda_0$ in Å = 1548.19
$A_E$ is the abundance relative to H (taken from G&L Table III): $3.2 \times 10^{-4}$
$q_i(w) = n_i/n_E$ is the ionization fraction of ion i of element E. Taken to be $10^{-4}$ based on the low density of C IV ions in the wind found from the BSEILB fitting to the C IV profile and the calculation in section 7.

The calculation of $\dot{M}$ from the above equation is a long, tedious and intense one, that when done manually, can lead to many errors. The actual numerical calculations were done using a computer program. More details are presented in the Appendix. The result is :

$$\dot{M} = 2.51 \times 10^{-6} \ M_\odot /yr.$$

The uncertainty here is difficult to determine because of uncertainties of the numerous parameters that go into the calculation. One of the reasons for this rather large value for $\dot{M}$ may be the small value assumed for $q_i$. But this was done considering the fact that the fit to the C IV profile indicates a low density of C IV in the wind.

Another estimate of the mass loss rate was calculated using equation (12) from Vink, J. S, *et al*. (2000), which is presented there as a sequence of logarithmic summations.

Log $\dot{M}$ = - 6.697+2.194log(Ls/10⁵) =  -6.653
         - 1.313log(Ms/30.0)         = +0.3239
         - 1.226log(Vr/2.0)           = -0.1102
         + 0.933log(T$_{eff}$/40000.0)   = -0.03159
         -10.92log(T$_{eff}$/40000.0))²  = +0.7395

Adding this sequence results in Log $\dot{M}$ = -5.731, and hence,

$$\dot{M} = 1.86 \times 10^{-6} M_\odot / yr$$

In the above statements:

> Ls is the Luminosity of the primary star taken to be $10^{5.02}$ $L_\odot$ as given in Penny *et al*., 2008.
> Ms is the mass of the primary star taken to be 17 $M_\odot$ as given in Penny *et al*., 2008.
> Vr = $v_\infty/v_{esc}$ = 2335/949 = 2.46, where $v_{esc}$ is for a distance of $R_\star$ = 7.2 $R_\odot$.
> $T_{eff}$ of the primary taken as 37,000K as listed in Penny *et al*., 2008.

The above value of $\dot{M}$ is in remarkable agreement with that calculated from equation (12) of G&L. It is also interesting that the Vink *et al*. equation does not include a term for the ionization fraction, $q_i$, and is based on statistics for hot, single stars. Because of the latter fact, it may not be applicable to contact binary star systems. The results from G&L may be more correct, since it is calculated using specific parameters for the system coming from the BSEILB fitting of the C IV profile, though the difference is not significant.

Mass Loss rates, $\dot{M}q_i$, for similar stars are listed in Table VII of G&L with values ranging from $5.01 \times 10^{-9}$ to $3.98 \times 10^{-10}$. The value of the mass loss rate for EM Car was found to be $\dot{M}q_i = 2.7 \times 10^{-9}$ by Pfeiffer & Stickland (2004). But, none of these values specify a value for the ionization fraction. Therefore, it is difficult to compare these results with the values found in this paper using the G&L or Vink equations. However, mass loss rates between $10^{-7}$ and $10^{-6}$ for OB binary stars are not uncommon (Stevens *et al*., 1992).

### 9. **Discussion of Results**

The best fits to the *IUE* data indicate that both stars have relatively shallow wind envelopes with densities that decrease rapidly with distance from the stars. The profile fit at phase 0.533 is very different than the one at phase 0.229 indicating a dramatic change in the geometry and density of the wind. The parameter W1 = 1 in Tables 1 and 2 means all the absorption of the photospheric flux by the wind effectively takes place within 1 stellar radius from the photospheres. The fits were very sensitive to this parameter. A value of W1 = 1.5 doubled the wind emission and resulted in a P Cygni profile, which does not exist. It also extended the absorption profile to shorter wavelengths than what is seen.

The terminal velocity of the wind at orbital phase 0.533 occurs at 12.05 Å from the rest wavelength of the blue component of the C IV doublet. This translates to a terminal velocity of the wind to be 2335 km/sec., whereas the terminal velocity at phase 0.229 was 0.67 of that value. It is not clear why there is a difference but it must be related to the difference in aspect of the wind as seen at these phases.

There is also a notable difference in the strength of the photospheric absorption lines that is also related to variable clouds in the winds or wrap around wind streams that effectively change the reversing layers of the stars.

The $W_{Gauss}$ parameter indicates some amount of turbulence in the wind flow, especially near the contact point of the stars. Including a larger amount of turbulence in the wind flow was found to produce a fit that was not as good. In effect, it causes the sharp decrease seen in the absorption near terminal velocity to be pushed to shorter wavelengths.

A geometrically separate shock between the stars cannot be detected in the spectrum, as was detected in EM Car (Pfeiffer *et al*., 2004). This is probably the result of the overcontact nature of the stars. The entire region of the common wind envelope between the stars is like a shock front and is very lumpy including various clouds and an emission cloud seen at phase 0.229 that is not seen in any other image. A study of the changing metallicity of the system as revealed in Fig. 5 is further testimony of this.

The apparent dearth of C IV in the winds must be a result of the high degree of ionization of C ions. The ionization potential of C IV to C V is 64.5 ev. From kinetic theory, the mean kinetic energy of all particles in the wind would have this value for T = 500,000 K. The average speed of the wind at a distance of 0.5 stellar radius is about 1,000 km/sec., since the terminal velocity of the wind at 1 stellar radius has been determined to be about 2300 km/sec. Assuming the wind mainly consists of H and He ions in the ratio of 7 to 3, and if their speed is thermalized due to turbulence in the wind, then T = $mv^2/3k$ = 765,000 K.

But the temperature in the wind does not need to be this high to ionize C IV to higher states of ionization. Even at T = 39,000 K, the Maxwell-Boltzmann kinetic energy distribution function indicates that, in the high energy tail of the distribution, there are some particles with this energy. One must also consider the very short wavelength radiation from the photospheres of the stars that baths the winds. For a surface temperature of 38,700 K for the primary, taken from Penney et al, (2008), Wien's Law gives a value of 749 Å for $\lambda_{max}$. Also, all photons carrying 64.5 ev or more corresponds to a wavelength of 64.1 Å, which is just inside the X-ray region of the EM spectrum. From the black body curve for 38,700 K there should be copious amounts of photons flowing into the winds to result in a high amount of ionization for most ions and, therefore, a high electron density.

So, it is a wonder that there is any C IV present to produce the absorption seen in the C IV profile. However, the Saha ionization equation does indicate that a high electron density environment in the winds could result in a recombination rate that could salvage an amount of C IV necessary to produce the absorption seen in the profiles.

A study of the line blanketing or metallicity of the stars reveals changes that are somewhat erratic. There are sudden jumps both with orbital phase and time, indicating that the winds are lumpy, consisting of various clouds that may come and go and/or eddies in a turbulent shock front. This is in agreement with the hydrodynamic calculations of Stevens *et al.*, (1992) for colliding winds in OB binary stars.

The co-rotational velocity of the outer limb of the primary star at primary conjunction was determined to be 298 km/sec., and to be near 500 km/sec at first quadrature. However, the concept of individual rotational velocities for the stars in an over-contact, co-rotational binary system is a problematic one.

The mass loss rate for the system is somewhat large, being in the neighborhood of $10^{-6}$ M$_\odot$/yr. as calculated from two independent relationships. However, such mass loss rates are common for O stars (Stevens *et al*., 1992). This may also be related to the fact that we have two stars whereas most published values for mass loss rates are for single stars without a value given for the ionization fraction.

### 10. Acknowledgements

My colleague, the late Robert H. Koch, has written most of the discussion on the mass ratio controversy of the TU Muscae System. I would also like to thank H. J. G. L. Lamers for providing Bob Koch and me with a copy of his SEIDOUB3.FOR program.

# 11. References

Andersen, J. and Clausen, J. V. 1989, A&A, **213**, 183.
Binnendijk, L, Properties of Double Stars, Univ. of Penn. Press, Philadelphia, 1960.
Bradstreet, D.H. (1993). Binary Maker 2.0 - An Interactive Graphical Tool for Preliminary Light Curve Analysis "Double Stars for the Masses". In: Milone, E.F. (eds) Light Curve Modeling of Eclipsing Binary Stars. Springer, New York, NY.
Groenewegen, M. A. T. & Lamers, H. J. G. L. M. 1989, A&AS, **79,** 359, 1989.
Howarth, I. D. & Prinja, R. K. 1989, ApJS., **69**, 527.
Kallrath, J. & Milone, E. F. 1999, *Eclipsing Binary Stars:Modeling and Analysis,* Springer- Verlag, New York.
Koch, R. H. 2001, Private Communication.
Kurucz, R. L. 1994, CD-ROM 19, Solar Abundance Model Atmospheres for 0, 1, 2, 4, 8 km/s (Cambridge: SAO)
Lamers, H. J. G. L. M., Cerruti-Sola, M. & Perinoto, M. 1987, ApJ, **314**, 726.
Lamers, H. J. G. L. M, Haser, S., De Koter, A. & Leitherer, C. 1999 ApJ, **516**, 872.
Pachoulakis, I. 1996, MNRAS, **280**,153.
Penny, L. R., Ouzts, C & Gies, D. R. 2008, Ap. J., **681**, 554.
Pfeiffer, R. J., Pachoulakis, I, Koch, R. H. & Stickland, D. J. 1994, The Observatory, **114**, 2
Pfeiffer, R. J. & Stickland, D. J. 2004, The Observatory, **124,** 117.
Stevens, I. R., Blondin, J. M. & Pollock, A. M. T. 1992, ApJ., **386,** 265.
Stickland, D. J., Lloyd, C., Koch, R. H. & Pachoulakis, I. 1995, *The Observatory,* **115**, 317.
Stickland, D. J. 1994, The Observatory, **114**, 297.
Terrell, D., Munari, U., Zwitter, T. & Nelson, R. H. 2003, AJ, **126**, 2998.
Van Hamme, W. 1993, AJ, **106**, 2096.
Vink, J. S., de Koter, A. & Lamers, H. J. G. I. 2000, A&A, **362**, 295.
Wilson, R. E. & Devinney, E. J. 1971, ApJ, **166**, 605.
Wilson, R. E. & Refert, J. B. 1981, ApSpSc, **76**, 23.

# 12. Appendix

Regarding the calculation of the mass loss rate: From equation (3) in G&L:

$w = w_0 + (1-w_0)(1-1/x)^\beta$, where $w_0$ is 0.01, $x = 1.5$ & beta = 0.50.

So, $dw/dx = d[w_0 + (1-w_0)(1-1/x)^\beta]/dx$
$= (1 - w_0)\beta(1- 1/x)^{(\beta-1)}(1/x^2)$
$= (1- 0.01)(0.5)(1-1/1.5)^{(0.5-1)}(1/1.5^2)$
$= 1.21$

From equation (5) in G&L: $\tau(w)=(T/I)(w/w_1)^{a_0}\{1-(w/w_1)^{1/\beta}\}^{a_1}$,
where the value for T/I emerges from the *BSEILB* program to be 8.897, a0 = 0.05, & a1 = 7.0. Then:

$w/w_1 = 0.5$, since $w = 0.5$ and $w_1 = 1$.
$(w/w_1)^{a_0} = 0.5^{0.05} = 0.966$
$(w/w_1)^{1/\beta} = 0.5^{1/0.5} = 0.5^2 = 0.25$
$1-(w/w_1)^{1/\beta} = (1-0.25) = 0.75$
$\{0.75\}^{7.0} = 0.133$

Therefore,   $\tau(w) = 9.395(0.966)(0.133) = 1.21$

Then:
$$\dot{M} = 1.189 \times 10^{-18} [R_* v_\infty^2 \tau(w) x^2 w(dw/dx)][f \lambda_0 A_E q_i(w)]^{-1}$$
$$= 1.189 \times 10^{-18} [7.2(2335)^2(1.21)(2.25)(0.5)(1.21)][0.194(1548.19)(3.2 \times 10^{-4})(10^{-4})]^{-1}$$
$$= 1.189 \times 10^{-18} [(3.93 \times 10^7)(0.517)]/[(300.3)(3.2 \times 10^{-8})]$$
$$= 1.189 \times 10^{-18} [2.03 \times 10^7]/[961 \times 10^{-8}]$$
$$\dot{M} = 2.51 \times 10^{-6}\ M_\odot/\text{yr}.$$